\documentclass[a4paper,12pt]{article}
\usepackage{epsfig}
\newcommand{\gs}   {$\gamma_{\rm s}\ $}

\begin{document}
\begin{center}
{\bf\Large{Status of Chemical Equilibrium\\
 in \\
Relativistic Heavy Ion Collisions}}
\end{center}
\begin{center}
\large{J. Cleymans}\\
{\it Department of Physics, University of Cape Town,\\
Rondebosch 7701, Cape Town, South Africa}\\
\end{center}
\begin{abstract}
Recent work on chemical equilibrium in heavy ion collisions 
is reviewed. The energy dependence of thermal parameters
is discussed. The centrality dependence of thermal
parameters at SPS energies is presented.
\end{abstract}
\section{Introduction}
\begin{figure}[htbp]
\begin{center}
\epsfig{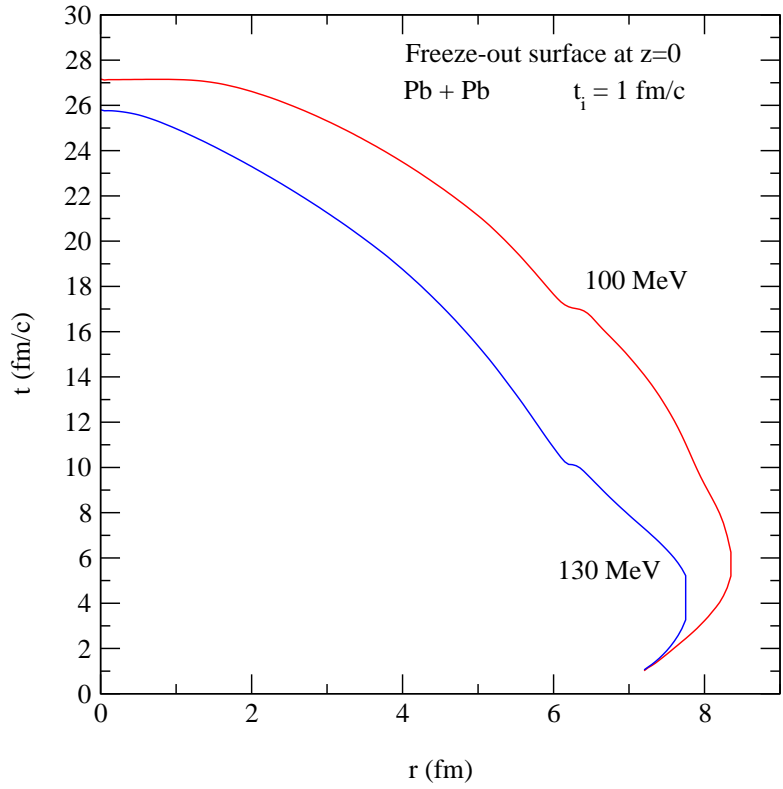}
\end{center}
\caption{Freeze-out surface obtained in Ref. 1. The
values for the  thermal freeze-out temperature are indicated.}
\label{fig:jaipur2_fig1}
\end{figure}

\begin{figure}[htbp]
\begin{center}
\epsfig{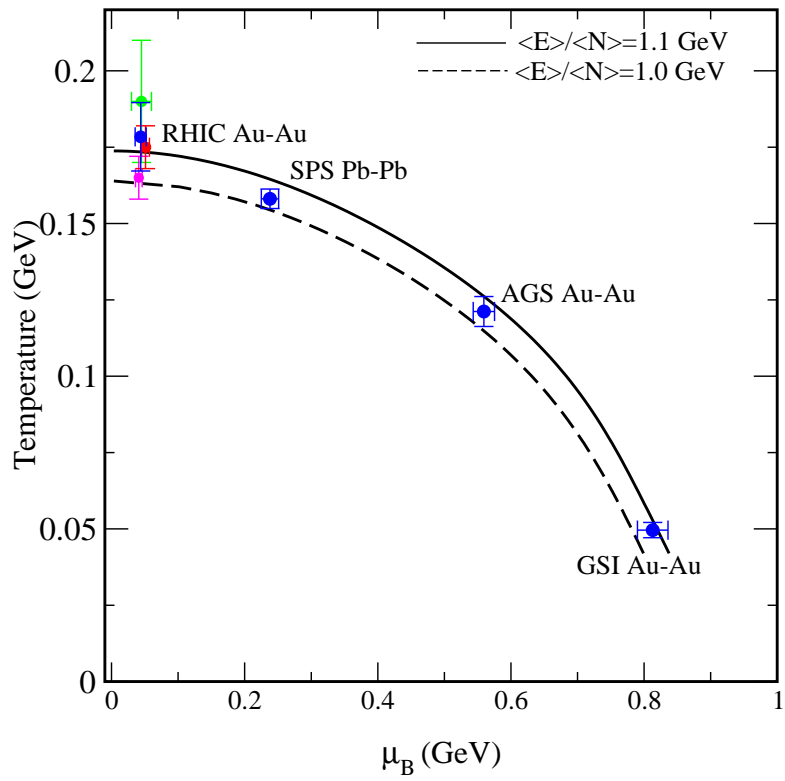}
\end{center}
\caption{Values of thermal parameters at
 different energies.}
\label{fig:jaipur2_fig2}
\end{figure}

In hydrodynamic models \cite{dinesh} the freeze-out surface 
is very sensitive on the initial conditions and is therefore
highly model dependent, (see e.g. Fig.~\ref{fig:jaipur2_fig1}).
However, this dependence disappears when considering 
 ratios of integrated particle 
yields, $N^i/N^j$, which are 
independent of the details of the initial conditions
and are the same as those of point-like particles 
in a fireball at rest, $N^i_0/N^j_0$ \cite{prc}. If 
the rapidity distribution is flat 
(as is the case at RHIC energies)  
then it is furthermore  possible to show that
\begin{equation}
\frac{\left.dN^i/dy\right|_{y=0}}{\left.dN^j/dy\right|_{y=0}}
=\frac{N^i_0}{N^j_0}  .
\end{equation}
The conditions for the validity of these results
have been discussed  in the literature \cite{prc,rischke,heinz}
and will not be repeated here. 
%
%
Motivated by the above results, a systematic study 
of fully integrated particle multiplicities
in central Au--Au and Pb--Pb collisions at 
beam momenta of 1.7$A$ GeV, 11.6$A$ GeV 
(Au--Au) and 158$A$ GeV (Pb--Pb) 
has been performed \cite{bcksr}.
The close similarity of the colliding systems makes it possible to study
heavy ion collisions under definite initial conditions
 over a wide range of beam energies. 
It can  be concluded that a thermal model description of particle 
multiplicities, with additional strangeness suppression, is
 possible for each energy.
The resulting temperature $T$ and chemical potential $\mu_B$
are shown in Fig. \ref{fig:jaipur2_fig2}
which also shows results from recent fits
 to the RHIC data \cite{xu,pbm3,poland}
It can been seen that, at all energies,  the average
 energy per hadron in the 
comoving frame is always close to 1 GeV per hadron
 despite the fact that the center-of-mass energy varies
more than 100-fold \cite{prl}:
\begin{equation}
{\left< E\right>\over\left< N\right>} \approx 1~\mathrm{GeV}.
\label{eqn:eovern}
\end{equation}
It should be noted that, with the advent of more precise data from
the NA49 collaboration,  this relation is 
now better satisfied than when it was 
first proposed in 1998 \cite{prl}.
%
\section{Energy Dependence of Thermal Parameters.}
\begin{figure}
\begin{center}
\epsfig{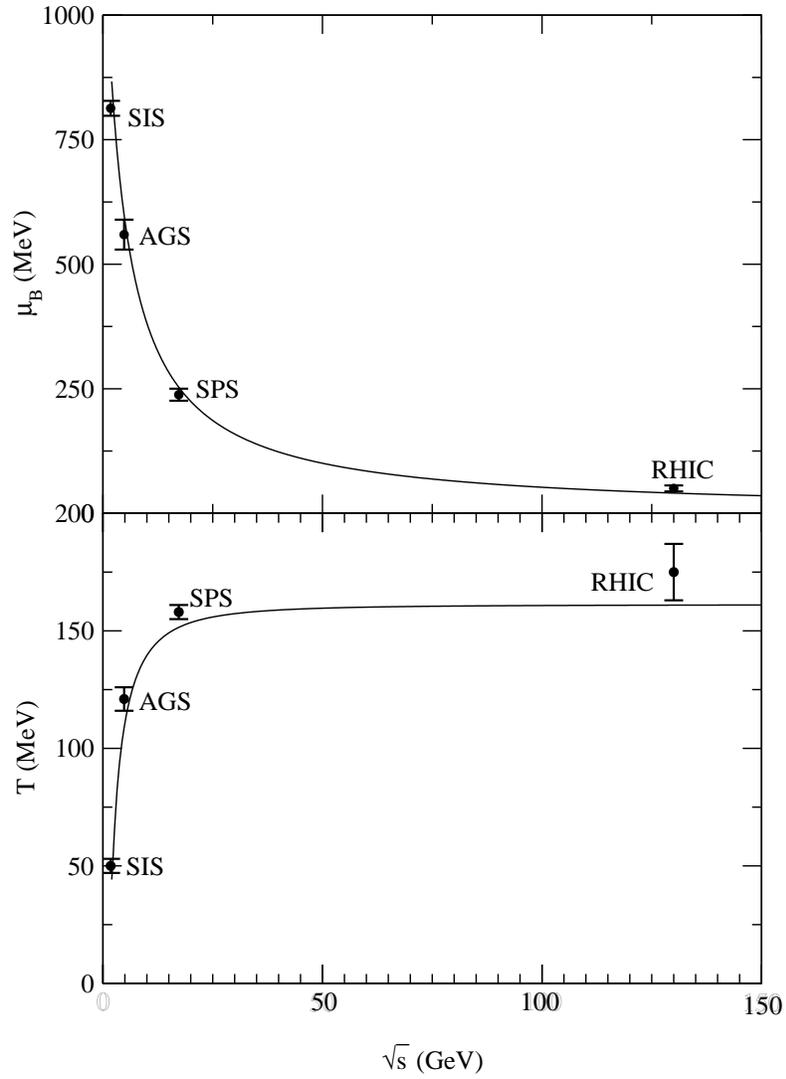}
\end{center}
\caption{Energy dependence of the temperature and the baryo-chemical potential.}
\label{fig:tmu}
\end{figure}

The energy dependence of the 
baryon chemical potential, $\mu_B$, can be described by a simple fit \cite{npa}
\begin{equation}
\mu_B(s) = {a\over 1 + \sqrt{s}/b}\
\end{equation}
with $a\approx 1.27$ GeV and $b\approx 4.3$ GeV.
The temperature is then determined from Eq.~\ref{eqn:eovern}.
The resulting fits to $\mu_B(s)$ and $T(s)$ are shown in Fig.~\ref{fig:tmu}.
This parametrisation can be used to 
discuss the energy dependence of particle ratios.
The temperature varies considerably between the lowest and the highest
 beam energies, 
namely, between 50 MeV at SIS and 160 MeV at SPS. Similarly, the baryon
 chemical potential changes appreciably, decreasing from 
about 820 MeV at SIS to about 240 
MeV at SPS. 
The supplementary \gs factor \cite{gs}, measuring
the deviation from a completely 
equilibrated hadron gas, is around 0.7 -- 0.8 at all energies where it has
been considered as a free fit parameter. 
At the present level of accuracy, 
a fully equilibrated hadron gas (i.e. \gs=1) cannot be ruled
 out in all examined
collisions except in Pb--Pb, where \gs deviates from 1 by more than $4\sigma$.
This result does not agree with a recent similar analysis of Pb--Pb data 
\cite{heppe} imposing  full strangeness equilibrium. The main reason for this
discrepancy is to be found in the different data set used; whilst in 
ref.~\cite{heppe} measurements in different limited rapidity intervals have
been collected, we have used only  particle yields extrapolated to 
full phase space. The temperature values that we have found essentially agree 
with previous analyses in Au--Au collisions \cite{cor} and 
estimates at 11.7 $A$ GeV \cite{stachel}.\\ 
\section{Maximum Relative Strangeness.}
%
The results of the previous section combined with the energy
dependence of the thermal parameters shown in  Fig.~\ref{fig:tmu} provide a
basis for the study of the energy dependence of strangeness production
in heavy ion collisions. Of particular interest are  the ratios of
strange to non-strange particle multiplicities as well as the
relative strangeness content of the system as expressed by the
Wroblewski factor \cite{wroblewski} defined as:
\begin{equation}
\lambda_s = {2\left<s\bar{s}\right>
\over \left<u\bar{u}\right> + \left<d\bar{d}\right>}
\end{equation}
Furthermore, we assume (for simplicity) strangeness 
to be in complete equilibrium,
that is, the  strangeness saturation factor is taken as  \gs=1.
When more data will become available, this point can be refined.

\begin{figure}
\begin{center}
\epsfig{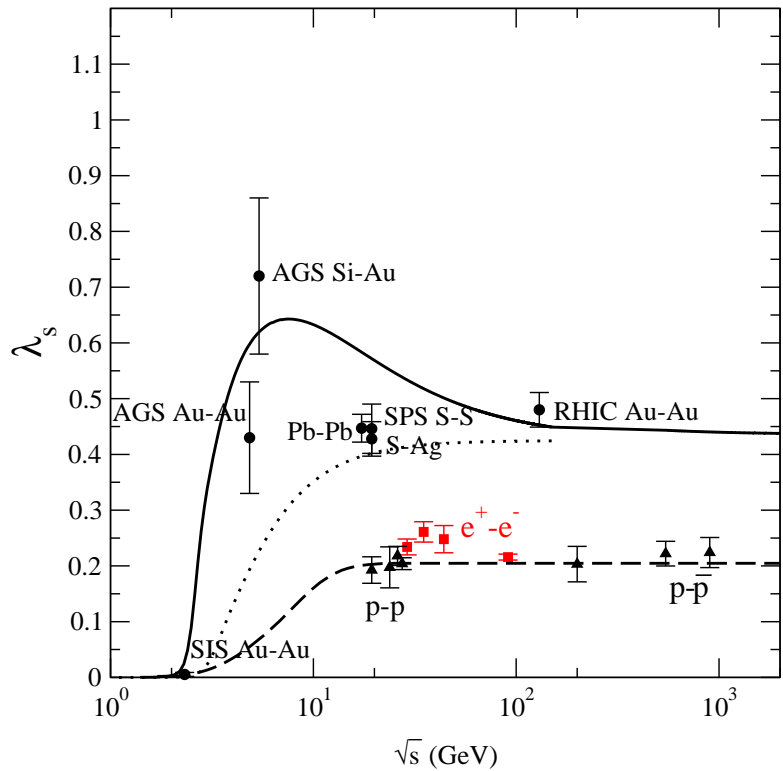}
\end{center}
\caption{Energy dependence of the Wroblewski factor. The full line
 has been calculated using the energy dependence 
of $T$ and $\mu_B$ given in Eq. (3) with
\gs = 1. The dotted line was calculated keeping $\mu_B = 0$, the
dashed line was calculated using the canonical ensemble 
with a radius fixed at 1.2 fm.}
\label{fig:jaipur2_fig4}
\end{figure}

We turn our attention first
to  the energy dependence of the Wroblewski ratio
The quark content used in this ratio is determined at the moment
 of chemical freeze-out, i.e.
from the hadrons and especially, hadronic
resonances, before they decay. 
This ratio is thus not an easily measurable  observable
unless one can reconstruct all resonances from the final-state
particles.
 The results are shown in Fig.~\ref{fig:jaipur2_fig4} as a function of
invariant energy $\sqrt{s}$. The values calculated from the
experimental data at chemical freeze-out  in central A-A
collisions have been taken from
reference~\cite{bcksr}.
There the statistical model
was fitted with the  extra parameter $\gamma_s$ to account for
the possible chemical under-saturation of strangeness. 
In general, values for $\gamma_s$ close to 1 are found.
At the SPS,
$\gamma_s\simeq 0.7$ gives the best agreement with 4$\pi$ data.
This is also the reason why the SPS points in Fig.~\ref{fig:jaipur2_fig4}
 lie below the
solid line.
The values of $\lambda_s$ were extracted from 4$\pi$ integrated
data with the exception of the result form RHIC where
particle ratios were measured only at mid-rapidity \cite{harris}.
The solid line in Fig.~\ref{fig:jaipur2_fig4} describes the statistical model
calculations in complete equilibrium along the unified freeze-out
curve~\cite{prl} and with the energy dependent thermal parameters
presented here. From Fig.~\ref{fig:jaipur2_fig4} we conclude
 that around 30 A$\cdot$GeV
lab energy the relative strangeness content in heavy ion
collisions reaches a
 clear and well pronounced maximum.
The Wroblewski factor  decreases towards higher incident energies
and reaches a value of about 0.43.

 The
appearance of the maximum can be traced  to the specific
dependence of $\mu_B$ on
the beam energy. 
In Fig.~\ref{fig:jaipur2_fig4}  we also show the results  for  $\lambda_s$ calculated
under the assumption that only the temperature  varies with
collision energy but the baryon chemical potential is kept fixed
at zero. In this case the Wroblewski factor  is indeed seen to be
a monotonic function of energy. The assumption of vanishing net
baryon density is of course the prevailing situation in e.g.
p-$\bar{\rm{p}}$ and e$^+$-e$^-$ collisions. 
In Fig.~\ref{fig:jaipur2_fig4} the
 results
for $\lambda_s$ extracted from the data in p-p, p-$\bar {\rm p}$ and
e$^+$-e$^-$ are also included~\cite{becattini}. The dashed line
represents results with $\mu_B= 0$ and a  radius of 1.2 fm. There
are two important differences in the behavior of $\lambda_s$ in
elementary compared to heavy ion collisions. Firstly, the
strangeness content is smaller by a factor of two.
In elementary collisions particle multiplicities follow
the values given by the canonical ensemble with radius 1.2 fm
 whereas in A-A collisions there is a transition from
canonical to grand canonical behavior. Secondly, there is no
significant maximum in the behavior of $\lambda_s$ in elementary
collisions
 due to the  zero baryon density in the p-$\bar{\rm{p}}$ and
e$^+$-e$^-$ systems.

Following the chemical freeze-out curve, shown as a thick
full line in
Fig.~\ref{fig:jaipur2_fig4}, one can see that
 $\lambda_s$ rises quickly from SIS to AGS energies,
then reaches  a maximum around $\mu_B\approx 500$ MeV
and $T\approx 130$ MeV.
These freeze-out parameters correspond to
30 GeV lab energy. At higher incident
energies the increase in $T$ becomes negligible but $\mu_B$ keeps
on decreasing and as a consequence $\lambda_s$ also decreases.

\begin{figure}
\begin{center}
\epsfig{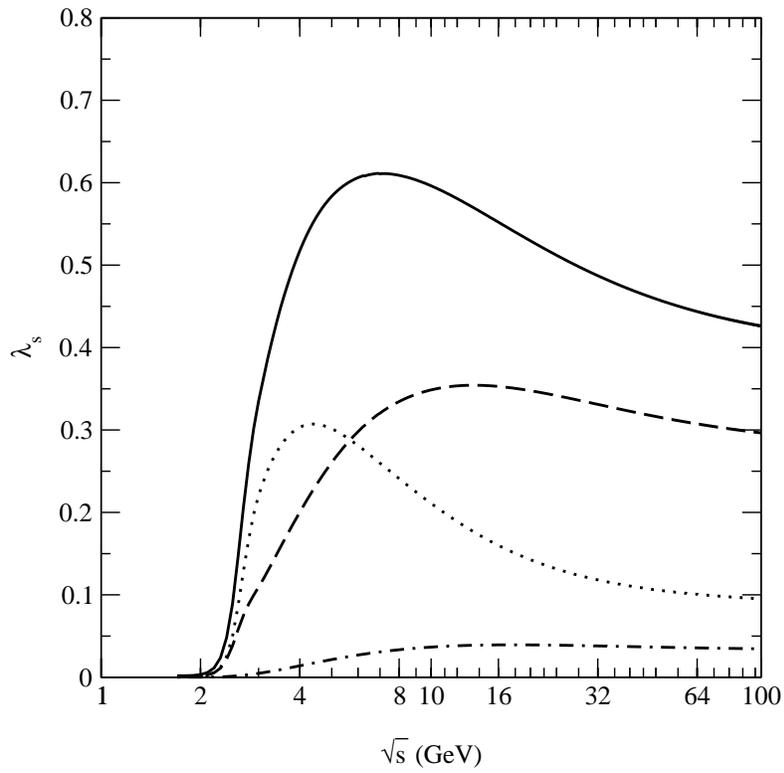}
\end{center}
\caption{Contributions to the  Wroblewski factor coming from
strange baryons (dotted line), strange mesons (dashed line)
 and hidden strangeness (dash-dotted line).}
\label{fig:jaipur2_fig5}
\end{figure}

\begin{figure}
\begin{center}
\epsfig{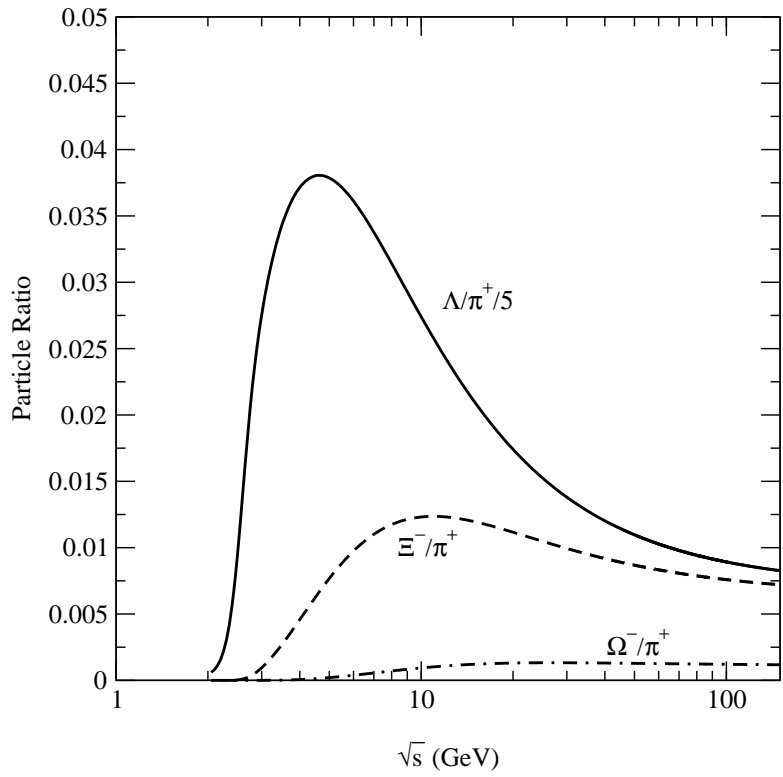}
\end{center}
\caption{\label{fig:jaipur2_fig6}Ratios of strange baryons to pions as a function
of energy for  \gs =1.}
\end{figure}

 The importance of finite baryon density on the
behavior of $\lambda_s$ is demonstrated in  Fig.~4  showing
 separately the  contributions to $\left<s\bar{s}\right>$
coming from strange
baryons, from strange mesons and from hidden strangeness, i.e.,
from hadrons  like $\phi$ and $\eta$.
 As can be seen in Fig.~\ref{fig:jaipur2_fig5},
the origin of the maximum in the Wroblewski
 ratio can be traced  to the contribution
of strange baryons.
Even strange mesons exhibit a broad maximum. This is due to the
presence of associated production of e.g.~kaons together with
hyperons. This channel dominates at low $\sqrt{s}$ and loses
importance at  high incident energies.
Next we study how the behavior of the Wroblewski ratio is reflected in
specific particle yields.
 The energy dependence of the
$\Lambda/\pi^+$ ratio is shown in Fig.~\ref{fig:jaipur2_fig6}.
The model  gives a good description of the data, showing
a broad maximum at the same energy as the one
seen in the Wroblewski factor.
The appearance of the maximum in the relative strangeness contribution
becomes also obvious when considering ratios which are more
sensitive to the baryon chemical potential. Fig.~\ref{fig:jaipur2_fig6}
also shows the energy
dependence of the $\Xi^- /\pi^+$ and
the $\Omega^-/\pi^+$ ratios  which all exhibit maxima.
 The  peak in the $\Lambda/\pi^+$ ratio
is  much stronger  than 
the one in the $\Xi^-/\pi^+$ or in the $\Omega^-/\pi^+$ ratios.
There is also a shift of the maximum  to higher energies for particles with
increasing strangeness quantum number.
These  differences appear  as a consequence  of the
enhanced strangeness content of the particles which suppresses
the dependence of the corresponding ratio on $\mu_B$.
\section{Centrality Dependence at SPS.}
We next analyze the centrality dependence of the thermal
parameters describing hadron multiplicities \cite{ckw}.
This  provides further information 
about the effects of the size of the excited strongly interacting
system and help in the systematic understanding of the experimental data. 
We will show that the thermal model is able to
describe the available data for various centrality classes at one beam energy.

In order to have a sound basis for the application of the 
thermal model we rely as much as possible
on fully integrated particle multiplicities. For this reason we 
concentrate our efforts on the analysis of  results obtained by 
the NA49 collaboration \cite{Sikler} using 
centrality selected fixed-target Pb--Pb collisions
at a beam energy of 158 GeV per nucleon, which are here analyzed
within the framework of the thermal model.

\begin{figure}
\begin{center}
\epsfig{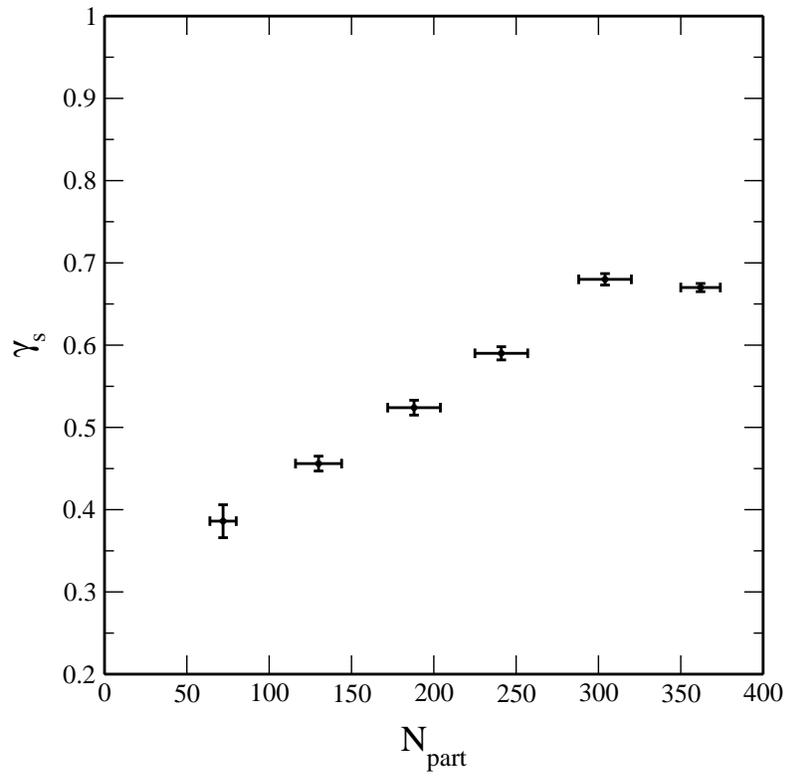}
\end{center}
\caption{\label{fig:jaipur2_fig7}Centrality dependence of the 
strangeness suppression factor \gs.}
\end{figure}

As shown in \cite{bcksr},
the inclusion or omission of certain hadron species can change 
considerably the extracted values of $T$ and $\mu_B$. We 
stress however that our analysis, due to the restricted available
data, focuses on the trends with changing centrality.
The temperature, $T$, and the baryon chemical potential, $\mu_B$,
do not show any noticeable  dependence on centrality \cite{ckw}.
%
There are two thermal parameters which exhibit a pronounced
dependence on  centrality: the strangeness equilibration
factor $\gamma_s$ increases 
approximately linearly and saturates for increasing centrality,
as shown in Fig.~\ref{fig:jaipur2_fig7}.
Other thermodynamic state variables,
such as the energy per hadron $\langle E \rangle / \langle N \rangle$,
the energy density, the baryon density, and the entropy per baryon
$S/B$,  remain fairly independent of the centrality. 

In summary,
the analysis of the thermal parameters, describing the 
 integrated yields of $\pi^{\pm}$, $K^{\pm}$, and $\bar p$
as obtained by the NA49 experiment \cite{Sikler} 
shows that  
the radius of the fireball increases linearly
with increasing centrality. Also
the strangeness parameter $\gamma_s$ increases,
i.e., strange particle multiplicities approach  chemical equilibrium.
In contrast, the temperature 
and the baryon chemical potential 
do not change  with centrality.
No in-medium modifications are needed to describe the above quoted
hadron yields. 
\section*{Acknowledgments}
Helpful discussions 
with H. Oeschler, K. Redlich, F. Becattini, A. Ker\"anen, E. Suhonen,
 B. Kaempfer, P. Braun-Munzinger, J. Stachel,
S. Wheaton, S. Yacoob, M. Marais
are acknowledged. Thanks are also due to D. R\"ohrich, P. Seyboth and R. Stock 
for invaluable help with the NA49 data.
%

%
\newpage
%
%
\end{document}